\documentclass[useAMS,usenatbib]{mn2e}

\usepackage{graphicx}
\usepackage{multirow}
\usepackage{natbib}
\usepackage{amsmath}
\usepackage{soul}

\newcommand{\msun}{M_{\odot}}
\newcommand{\rsun}{R_{\odot}}

\def\mpy{{\rm ~M}_{\odot} {\rm ~yr}^{-1}}

\title[The formation and evolution of MWC~656.]
      {On the formation and evolution \\of the first B\lowercase{e} star in~a~black~hole binary MWC~656}
\author[M. Grudzinska et al.]
       {M. Grudzinska,$^1$\thanks{E-mail: mgrudzinska@astrouw.edu.pl} 
        K. Belczynski,$^1$\thanks{Warsaw Virgo Group} 
        J. Casares,$^{2,3}$
        S.E. de Mink,$^4$ 
        J. Ziolkowski,$^5$ 
        \newauthor
        I. Negueruela,$^6$ 
        M. Rib\'o,$^7$\thanks{Serra H\'unter Fellow} 
        I. Ribas,$^8$ 
        J.M. Paredes,$^7$ 
        A. Herrero,$^{2,3}$ 
        M. Benacquista$^9$ \\
       $^1$ Astronomical Observatory, University of Warsaw, 
            Al. Ujazdowskie 4, 00-478 Warsaw, Poland \\
       $^2$ Instituto de Astrofisica de Canarias, 
            E-38205 La Laguna, S/C de Tenerife, Spain \\
       $^3$ Departamento de Astrofisica, Universidad de La Laguna,
            E-38206 La Laguna, S/C de Tenerife, Spain\\
       $^4$ Anton Pannekoek Astronomical Institute, University of Amsterdam, 
            1098 HX Amsterdam, The Netherlands\\
       $^5$ Nicolaus Copernicus Astronomical Center, 
            Bartycka 18, 00-716 Warsaw, Poland \\
       $^6$ Departamento de Fisica, Ingenieria de Sistemas y Teoria de la 
            Senal, Universidad de Alicante, Apartado, 99, \\ E-03080, Alicante, 
            Spain\\
       $^7$ Departament d'Astronomia i Meteorologia, Institut de Ci\`encies del
            Cosmos, Universitat de Barcelona, IEEC-UB, \\ Mart\'{\i} i Franqu\`es 1,
            E-08028 Barcelona, Spain \\
       $^8$ Institut de Ciencies de l'Espai - (IEEC--CSIC), Campus UAB, 
            Facultat de Ciencies, Torre C5 -- parell -- 2a planta, \\
            E-08193 Bellaterra, Spain \\
       $^9$ Center of Gravitational Wave Astronomy, University of Texas 
            at Brownsville, Brownsville, TX 78520, USA 
}

\date{\today}

\begin{document}

\maketitle

\begin{abstract}

We find that the formation of MWC 656 (the first Be binary containing a black 
hole) involves a common envelope phase and a supernova explosion. This result 
supports the idea that a rapidly rotating Be star can emerge out of a common 
envelope phase, which is very intriguing because this evolutionary stage is 
thought to be too fast to lead to significant accretion and spin up of the B 
star.
We predict $\sim 10$--$100$ of B-BH binaries to currently reside in the 
Galactic disk, among which around $1/3$ contain a Be star,
but there is only a small chance to observe a system with 
parameters resembling MWC~656. If MWC~656 is representative of intrinsic 
Galactic Be-BH binary population, it may indicate that standard evolutionary 
theory needs to be revised. This would pose another evolutionary problem in 
understanding BH binaries, with BH X-ray Novae formation issue being the 
prime example. 
The future evolution of MWC~656 with a $\sim 5\msun$ black hole and with a 
$\sim 13\msun$ main sequence companion on a $\sim 60$ day orbit may lead to 
the formation of a coalescing BH--NS system. The estimated Advanced LIGO/Virgo 
detection rate of such systems is up to $\sim 0.2$ yr$^{-1}$. This empirical 
estimate is a lower limit as it is obtained with only one particular 
evolutionary scenario, the MWC~656 binary. This is only a third such 
estimate available (after Cyg X-1 and Cyg X-3), and it lends additional 
support to the existence of so far undetected BH--NS binaries.

\end{abstract}


\section{Introduction}

We know at present 184 X-ray binaries consisting of a Be star and a
compact object - Be~XRBs \citep{Ziolkowski2014}. Until the previous
year, whenever the nature of the compact component was determined
(in 119 systems), it was always a neutron star. Not a single Be system 
containing a black hole was found during 40 years of observations of 
Be~XRBs. Last year the first such system was finally found by 
\cite{Casares_2014}. This discovery motivated us to investigate the 
possible evolutionary scenarios leading to the formation of similar 
systems.

\subsection{Be stars}

Be stars are massive, not substantially evolved, main sequence stars
of spectral types B0-A0 with Balmer emission lines
\citep{PorterRivinius2003}. This range of spectral types corresponds
roughly to a mass range of about 3 to $18 \msun$. The emission lines
(which give the name to this class of stars) originate in an
outflowing viscous disc (excretion disc) around the star. Such a
disc is very similar to the well known viscous accretion discs,
except for the changed sign of the rate of the mass flow. The
excretion discs evolve dynamically on a time scale of a few years to
few decades. In the course of this evolution, the disc undergoes a
global one-armed oscillation instability \citep{Kato1983},
manifesting itself in the form of the well observed, so called, V/R
variability. This instability (progressing density waves) leads
eventually to the disruption of the disc. A disc-less phase (with no
emission lines) follows then, until the disc refills again (which
takes years to decades \citep{PorterRivinius2003}). The excretion
discs were successfully modeled \citep{Hummel1995, Okazaki1996,
Hummel1997, Okazaki1997, Porter1999, NegueruelaOkazaki2000,
NegueruelaOkazaki2001} and these modelings helped to explain many
observed properties of Be stars (both solitary ones and those that
are members of binaries, in particular, X-ray binaries).

\subsection{Fraction of Be stars among B stars}
\label{Be_among_B}

At first, we would like to comment on the notion that the same star
might show a Be phenomenon (and so be a Be star) over some intervals
of time and not to show it (and so be a ``normal" B star) over some
other intervals. It is quite likely, that a ``normal" B star in the 
course of its evolution is spun up and develops a Be phenomenon. 
However, it is clear that there are B stars (majority of them) that 
{\it are not} Be stars and that there are Be stars that show Be 
phenomenon during all time we observed them. They exhibit disc-less 
phases, which may last decades \citep{PorterRivinius2003}. Such 
phases are part of a Be phenomenon and the stars do not stop being 
Be stars (at least, this is the accepted convention).

After this clarification, let us estimate how large a fraction of all B stars
are Be stars. \cite{Abt1987} found that Be stars comprise $18\%$ of the B0-B7
stars in a volume-limited sample of field stars (with a maximum Be fraction
for the spectral types B3-B4). A similar result was obtained by \cite{Zorec1997}
who found the fraction of Be stars to be $17\%$ for the Galactic field stars
(this was a mean value; a maximum of $34\%$ was found for the spectral type
B1). The percentage of Be stars was estimated also for stellar clusters.
\cite{Keller1999} investigated the frequency of Be stars in six young clusters
in the Magellanic Clouds. They found a range of 13 to $34\%$. \cite{Maeder1999}
investigated 21 clusters in the interior of the Galaxy, the exterior of the
Galaxy, the LMC and the SMC and found the fractions of, respectively, 11, 19,
23 and $39\%$. The similar investigation carried out by \cite{WisnBjork2006}
brought similar results (the ranges of Be fractions were 9 to $39\%$ for
earlier type (B0-B3) Be stars and 3 to $32\%$ for later (B4-B5) types).
\cite{Swain2005} analyzed 48 open clusters and found the mean (for all
clusters) percentage of Be stars equal to $7.1\%$ (with a maximum of $\sim 11\%$
for spectral types B2-B3). They compare it with Abt's value and attribute
the difference to the selection effects in Abt's estimate (a bias towards
earlier B spectral types). \cite{Fabregat2000} analyzed seven ``Be rich"
clusters in Milky Way and in the Magellanic Clouds. They found very high
percentage: 21 to $\sim 50\%$. Another ``Be rich" cluster was investigated by
\cite{Marco2013} who found very high ($\sim 40\%$) fraction of Be stars close
to the turn-off (spectral type B1) but very few Be stars for later spectral
types.

To summarize, the fraction of Be stars among B stars is about 20-$30\%$,
generally increasing for earlier spectral types.
If we consider Be X-ray binaries, one should remember that Be stars in X-ray 
binaries have somewhat earlier spectral types than solitary Be stars (see the 
further text), which may indicate that the factor $f_{Be}$ (the fraction 
of X-ray binary containing a Be star) should be somewhat higher - perhaps $30\%$.

\subsection{The origin of fast rotation of Be stars}

There is little doubt that all distinct properties of Be stars are related 
to the presence of an outflowing excretion disc. There is also little 
doubt that the presence of these discs is related to the fast rotation of 
these stars. \cite{Struve1931} suggested that this rotation is very close to 
the critical (or break-up) equatorial velocity. Later, a canonical view was 
established \citep{Porter1996, Chauville2001} according to which the rotation
is significantly subcritical with equatorial velocity equal only 70-$80\%$ of 
the critical velocity. However, more recently \cite{Townsend2004} and 
\cite{Ekstr2008} gave arguments indicating that the rotation might be indeed 
very close to critical, with equatorial velocity smaller only by a few percent
(and not 20-$30\%$) than the critical one. Such fast velocity makes the 
formation and maintaining of the excretion disc much more likely 
\citep{Granada2013}.

As for the origin of fast rotation, \cite{Martayan2006} indicated that Be 
stars are born with higher initial (on ZAMS) rotation than other B stars. It 
seems, however, that they are not born as Be stars from the very beginning. 
Rather, the higher initial rotation facilitates the action of mechanisms that 
later spin up these stars to nearly critical rotation. Two such major 
mechanisms were considered. 

One of them is the evolutionary spin up during the Main Sequence (MS) evolution.
The reason for the spin up is the significant decreasing of the moment of 
inertia of the star during this phase of evolution. This explanation might be 
supported by the fact the Be phenomenon seems to be associated with the second 
half of the MS evolution of B stars \citep{Swain2005, Fremat2006}. The first 
mechanism was discussed and modeled by different authors \citep{Meynet2005, 
Ekstr2008, Granada2013}. The general conclusion of these modelings is that 
evolutionary spin up is sufficient to explain the Be phenomenon. 

The second mechanism is the spin up due to accretion in a binary system. 
Initially this scenario was proposed for the formation of Be-X-ray binaries 
(e.g., \cite{Rappaport1982}). This scenario can also account for single Be 
stars, due to disrupted binary systems or binary mergers. First estimates of 
the importance of this scenario were provided by \cite{Waters1989, Pols1991, 
Portegies1995}. More recent simulations accounting for the actual spin up 
process and mergers were performed by \cite{Mink2013}. The general view of the 
advocates of the binary mechanism is that this scenario can account for the 
majority of the Be stars. For example \cite{Swain2005} concluded that more 
than $70\%$ of all Be stars had to be spun up in the process of accretion in 
binary systems. \cite{Mink2013} and \cite{ShaoLi2014} both conclude that all 
Be stars can be accounted for by the binary evolution. Especially, if mergers 
of short period contact binaries are taken into account. 

The summary of this controversy is difficult. It seems possible that both 
mechanisms are at work and that, at present, it is not possible to estimate 
reliably the relative importance of each formation channel.

\subsection{Be stars in X-ray binaries}

Be stars in X-ray binaries are, in many respects, similar to isolated 
Be stars. They have excretion discs which develop one-armed oscillations 
and display V/R variability. The dynamical evolution of their discs 
includes the disruption of the disc and the following disc-less phase.

However, there are also notable differences. The first concerns the spectral 
types. Be stars in X-ray binaries have, on average, earlier spectral types 
than isolated Be stars \citep{Negueruela1998}. The range is O9 to B3 as 
opposed to B0 to A0 for isolated Be stars. It indicates that also masses of 
Be stars in X-ray binaries are somewhat higher (perhaps 6 to 24 M$_\odot$ 
instead of 3 to 18 M$_\odot$). The second major difference concerns the 
excretion discs. The discs in X-ray binaries interact with the compact 
companions (almost always neutron stars). This interaction leads to the 
appearance and the growth in the certain locations in the disc of the 
resonances between Keplerian frequencies of the disc matter and the orbital 
frequency of the neutron star. These resonances lead to the tidal truncation 
of the disc \citep{Artymowicz1994, NegueruelaOkazaki2000, 
NegueruelaOkazaki2001, OkazakiNegueruela2001}. Tidal truncation makes the 
discs in Be X-ray binaries smaller and denser than the discs around isolated 
Be stars \citep{Reig1997, NegueruelaOkazaki2001}.

\subsection{Be X-ray binaries}
\label{SecBeXRB}

Be X-ray binaries (Be XRBs) are the most numerous class among high
mass X-ray binaries. We know at present 184 Be XRBs and only about
60 other high mass X-ray binaries (both Galactic and extragalactic
systems are included in this statistics; \cite{Ziolkowski2014}). 
In 119 of Be XRB systems the X-ray pulsations are observed, confirming
that the compact component must be a neutron star. The pulse periods
are in the range of 34 ms to $\sim$ 1400 s (\cite{ZiolkBelcz2011}). 
Until the previous year not a single Be system containing a black hole 
was found. The Be XRBs are rather wide systems (orbital periods in the 
range of $\sim 10-1180$ days; \cite{Ziolkowski2014}). The orbits are 
frequently eccentric. A compact component accretes from the excretion 
disc of a Be star (earlier known as the equatorial wind of a Be star).

The X-ray emission from Be XRBs (with a few exceptions) is of a
distinctly transient nature with rather short (days to weeks) active
phases separated by much longer (months to tens of years) quiescent
intervals (a typical flaring behavior). There are two types of
flares, which are classified as Type I outbursts (smaller and
roughly regularly repeating) and Type II outbursts (larger and
irregular). This classification was  first defined by
\cite{Stella1986}. Type I bursts are observed in systems with
highly eccentric orbits. They occur close to periastron passages of
a neutron star. They are repeating at intervals $\sim P_{\rm orb}$.
Type II bursts may occur at any orbital phase. They are correlated
with the disruption of the excretion disc around Be star. They
repeat on time scale of the dynamical evolution of the excretion
disc ($\sim$ few years to few tens of years). This recurrence time
scale is generally much longer than the orbital period
\citep{Negueruela2001}.

Be XRBs systems are known to contain two discs: an excretion disc
around the Be star and an accretion disc around the neutron star.
Both discs are temporary: the excretion disc disperses and refills
on time scales $\sim$ few years to few decades (dynamical evolution
of the disc \citep{PorterRivinius2003}), while the accretion disc
disperses and refills on time scales $\sim$ weeks to years (which is
related either to the orbital motion of a neutron star on an
eccentric orbit or to the disruption episodes of the excretion
disc). Formation of the accretion discs was analyzed by
\citep{Hayasaki2006} and \citep{Cheng2014}.

The more detailed description of the properties of Be~XRBs is given, e.g. in 
\cite{Negueruela2001, Ziolkowski2002, Bel+Ziolk, ZiolkBelcz2011, Reig2011} and
references therein.

The fact that we observe over one hundred neutron star Be XRBs and not a 
single black hole Be XRB, became known as the problem of the missing black 
hole Be XRBs. Trying to explain the reasons for which we do not observe black 
hole Be XRBs, \cite{Bel+Ziolk} carried out stellar population synthesis 
calculations aimed at estimating the ratio of neutron star to black hole Be 
XRBs, expected on the basis of the stellar evolution theory. The results of 
their calculations predict that for our Galaxy the expected ratio of Be X-ray 
binaries with neutron stars to the ones with black holes $F_{\rm NS/BH}$ 
should be, most likely, equal $\sim$ 54. Since we know 48 neutron star Be 
systems in the Galaxy, then it comes out that the expected number of black
hole systems should be just one. It seems that this system was just found
\citep{Casares_2014}.

\section{Modeling}

\subsection{The {\tt StarTrack} code}

We use the {\tt StarTrack} population synthesis code \citep{Belczynski2002,
Belczynski2008} to generate a population of binaries in the Galaxy. The code 
is based on revised formulas from \cite{Hurley2000}, among other with updated 
wind mass loss prescriptions, and calibrated tidal interactions, physical 
estimation of donor's binding energy ($\lambda$) and convection driven, 
neutrino enhanced supernova engines. A full description of the code can be 
found in the papers mentioned above. Here we concentrate only on these aspects 
that are important from the viewpoint of modeling used in this study. 

The initial parameters of systems simulated with the {\tt StarTrack} are 
randomly chosen from the following distributions:
\begin{itemize}
\item[-] three component broken power-law initial mass function (IMF) with 
         slope of $-1.3$ for initial mass $M_{\rm zams}=0.08$-$0.5\msun$, 
         $-2.2$ for $M_{\rm zams}=0.5$-$1\msun$, $-2.7$ for 
         $M_{\rm zams}=1$-$150\msun$ \citep{Kroupa2003} 
\item[-] flat mass ratio distribution in range $q=0-1$ \citep{Kobulnicky2006} 
\item[-] flat in the logarithmic distribution of initial binary separations 
         \citep{Abt1983} $\propto 1/a$ in range from a minimum value, to 
         prevent stars from filling their Roche lobes at zero-age, up to 
         $10^5\rsun$ 
\item[-] thermal-equilibrium distribution of eccentricities 
         \citep{Duquennoy1991} $\Xi(e)= 2e$ in a range 0-1. 
\end{itemize}
The binary fraction is assumed to be $50\%$ (i.e., $2/3$ of stars are in 
binaries). We note that recently measured initial distributions for O stars 
\citep{Sana2012} are different from those employed in our study. However, it 
was demonstrated that a change of distributions from the ones used here to 
the new ones does not significantly affect predictions for progenitors of 
double compact objects (de Mink \& Belczynski 2015, in prep). 

Among the most important physical mechanisms driving binary evolution is the 
common envelope (CE) phase. The CE is very efficient in creating close 
binaries. The outcome of this evolutionary phase can be described by the 
energy balance formula \citep{Webbink1984}:
\begin{equation}
  \alpha_{\rm CE} \bigg( \frac{GM_{\rm don,f}M_{\rm acc}}{2A_{\rm f}} - 
  \frac{GM_{\rm don,i}M_{\rm acc}}{2A_{\rm i}} \bigg) = 
  \frac{GM_{\rm don,i}M_{\rm don,env}}{\lambda R_{\rm don,lob}}
\label{CE}
\end{equation}
where $M_{\rm don}$ and $M_{\rm acc}$ are masses of donor and accretor, 
respectively; $M_{\rm don,env}$ is mass of the envelope of the accretor; $A$ 
is a binary separation; $R_{\rm don,lob}$ is a Roche lobe radius of the donor 
at the beginning of mass transfer. Index $i$/$f$ indicates the initial/final 
(before/after CE) value of a given quantity. The $\lambda$ parameter describes
the binding energy of the envelope of the donor and in the current version of 
code we use the ``Nanjing'' $\lambda$ (\cite{Xu2010}, \cite{Xu2010b}) with 
specific implementation into the {\tt StarTrack} code described in 
\cite{Dominik2012}. The $\alpha_{\rm CE}$ parameter describes the efficiency 
of the transfer of orbital energy into the envelope. We allow for large 
variation of this parameter. 

Another important phase in the past evolution of binaries is the 
core-collapse/supernova (SN) explosion. Due to its potential asymmetry, the 
new born compact object (NS or BH) may receive a natal kick. According to the 
observed velocities of radio pulsar \citep{Hobbs2005}, we choose the maximum 
kick velocity from the single Maxwellian distribution with $\sigma = 265$ 
km~s$^{-1}$. This value can be reduced depending on the fallback factor 
($f_{\rm fb}$), which describes percentage amount of matter ejected during SN 
and accreted back onto the compact object:
\begin{equation}
  V_k = V_{max} (1-f_{\rm fb}) 
\label{Vkick}
\end{equation}
This prescription is used for neutron stars (NS) and black holes (BH), but 
most of the former receive full natal kicks, with exception of electron 
capture SN (ECSN) for which we adopt no natal kicks at all. The full 
description of double compact object formation and rationale behind it is 
given by \cite{Fryer2012} and \cite{Belczynski2012}.

\subsection{The standard model}

In the standard model we employ energy balance for CE evolution with fully 
efficient transfer of orbital energy to the envelope energy ($\alpha=1$). The 
maximum natal kicks velocities are drawn from the Maxwellian distribution with
$\sigma=256$ km~s$^{-1}$. Fallback factor $f_{fb}$ varies in range 0-1 (from 
full kick for 0 to no kick for 1). The SN explosion mechanism for the standard
model is a convection driven, neutrino enhanced engine \citep{Fryer2012}. The 
explosion occurs within the first $0.1-0.2$ s (so called ''rapid'' explosion).
This engine reproduces \citep{Belczynski2012} the observed Galactic X-ray 
binary mass gap \citep{Ozel, Bailyn}. All results that we present are 
obtained for the metallicity typical of the Solar neighbourhood $Z=0.02$ and 
presented for a specific assumption on CE outcome: all donors beyond main 
sequence are allowed to survive CE (but see \cite{Belczynski2007,
Belczynski2010} for an alternative scenario for Hertzsprung Gap donors). Full 
description of the standard evolutionary model can be found in 
\cite{Dominik2012}. 

We only evolved binaries with primaries in mass range $6$-$150\msun$ and 
secondaries in mass range $1$-$150\msun$, as lower mass binaries may have only
a very little chance to produce binaries containing a BH. We only follow stars in the 
Galactic disk for which we assume a constant star formation at the level of 
$3.5\mpy$ over $10$ Gyr\footnotemark[1].
\footnotetext[1]{
\cite{Diehl_2006}, \cite{Mis2006}, \cite{Robitaille_2010} find a total 
star formation rate to be 4, 2.7, $\sim 1 \mpy$, respectively. A study 
by \cite{Kennicutt_1998} suggests SFR value in the range 0-10 $\mpy$.
Additionally, if we assume a constant SFR over the period of 10 Gyr then 
the Galactic disc mass provided by \cite{McMillan_2011} divided by this 
period gives an average SFR $\approx 6.5 \mpy$. Thus, the value of SFR 
chosen for our study ($3.5 \mpy$) is within reasonable limits.}
We have evolved $10^8$ binaries which gives us one full realization of the 
Galactic disk. We then use {\em Monte Carlo} techniques to evaluate the 
probability of catching each system during B-BH phase at the current time 
in the Galactic disk. A fraction $f_{Be}$ of B-BH binaries belong to 
the interesting group of Be-BH systems (see Sec.~\ref{Be_among_B}).

\subsection{Variations on the standard model}

We also check the sensitivity of our predictions to parameters that are 
important for the formation and evolution of B-BH binaries. 

In models $V_1$ and $V_2$ we change the $\alpha_{CE}$ value to $0.1$ and $5$ 
respectively. In model $V_3$ all BH do not receive any natal kicks during SN 
explosion. In model $V_4$ we change the mechanism of SN explosion from the 
``rapid'' explosion to the ``delayed'' one. In delayed SN engine, the 
explosion can occur as late as $1$s after bounce. This scenario, unlike the 
``rapid'' one, produces a continuous mass spectrum of compact objects, and 
the observed mass gap must then be a result of some observational bias 
\citep[e.g.,][]{Kreidberg2012}.

\subsection{Definition of B-BH and MWC~656 - like systems}

Due to large uncertainties in rotation physics we choose not to use 
rotation as a qualifier for the Be phenomenon. Initial rotation of massive 
stars is not yet fully constrained, angular momentum gain during accretion 
in massive binaries is not fully understood and angular momentum loss with
stellar winds for massive stars is still uncertain (see Sec.~\ref{discu} for a more
detailed discussion). Therefore, we study a broad spectrum of massive stars 
with BH, to which we refer as B-BH binaries and which are defined below. 
Some of these binaries may produce or are born with Be or Oe star.

In our simulation B-BH systems are binaries satisfying a following criterion:
\begin{equation}
\begin{gathered}
 BH + MS\ binary                \\
 10 < P_{\rm orb} < 1200{\rm d} \\
 3 < MS\ mass < 30 \msun.  
\end{gathered}
\label{BeBH}
\end{equation}
Note that this specific choice of orbital periods along with MS secondaries 
leads to population of wind-fed binaries. We have chosen the orbital periods 
such, that they are within the expected range for Be X-ray binaries (see 
Sec.~\ref{SecBeXRB}). Also note that MS mass range includes not only B 
stars ($<15\msun$) but also some O stars as to correspond to the observed 
Be/Oe population (\cite{Negueruela1998} and references therein).

We also define a subpopulation of B-BH binaries with properties similar to 
these observed for MWC~656 \cite{Casares_2014} as:
\begin{equation}
\begin{gathered}
  BH + MS\ binary \\ 
  55 < P_{\rm orb} < 65{\rm d}   \\
  3.8 < M_{BH} < 6.9\msun        \\
  10 < M_{MS} < 16\msun          \\
  e < 0.5
\end{gathered}
\label{MWC}
\end{equation}
The masses have been chosen within the one sigma limits determined by 
\cite{Casares_2014}; the eccentricity range was established in a private 
communication with the team which discovered the first BH+Be binary; the 
range for the orbital period was chosen arbitrarily. 

Note, that to obtain the information on Be-BH systems or systems similar to 
MWC~656 containing a Be star, all numbers/rates we get for the groups defined above 
need to be multiplied by a factor $f_{Be}$, which is not well known. If the 
fraction of systems where the B star shows the Be phenomenon were similar 
to the observed fraction in Milky Way and Magellanic Clouds clusters, the 
reduction factor $f_{Be}$ would be about 0.3 (see Sec.~\ref{Be_among_B}).

\section{Results}

In the following text we refer to initially more massive star as primary (BH 
progenitor), and initially less massive star (MS star that is potential Be or 
Oe object) as secondary.

\subsection{Overall properties of B-BH binaries}

The total number of B-BH systems formed in our standard model simulations over
entire $10$ Gyr of evolution of the Galactic disk is $N_{\rm B-BH}^{\rm form} 
= 8,700$, while number of MWC~656-like systems is $N_{\rm MWC656}^{\rm form} =
13$.

Note that our predictions for MWC~656-like systems are subject to errors 
from small number statistics. But these errors are smaller than 
ones associated with evolutionary uncertainties. Various systems that we 
generate have a range of lifetimes during B-BH (with an average of $\sim 45$ 
Myr) or MWC~656 (with an average of $\sim 6$ Myr) stage. We use the lifetime 
of each system to assess the probability that it is present in the current 
Galactic disk population. We use $N=10^4$ of Monte Carlo realizations of the 
formation time (drawn from uniform distribution) of a given system in range 
$0-10$ Gyr and check how many times $n$ a given system is present at the 
current Galactic age ($10$ Gyr). The current number of given binary population
is given by $n/N$. This number may be smaller than one and then it indicates 
how low is the probability of one system to exist in the current predicted 
population.  

The total number of B-BH systems found in our simulations to be present at the
current moment in the Galactic disk is $N_{\rm B-BH}^{\rm curr} = 39$, while 
number of MWC~656-like systems is well below one: $N_{\rm MWC656}^{\rm curr} =
0.007$ (i.e., the probability of having one system at present is $\sim 1\%$).

\begin{figure}
\begin{center}
\includegraphics[width=\columnwidth]{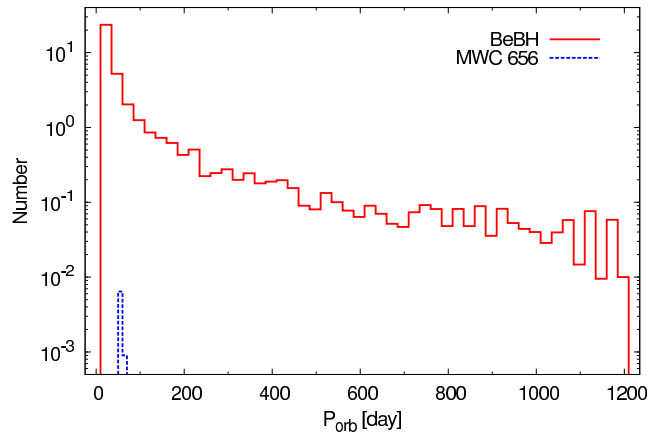}
\caption{Orbital period distribution for overall group of B-BH binaries 
(red solid line) and MWC~656-like subpopulation (blue dashed line). Predicted 
current Galactic populations are shown. Note that overall number of B-BH binaries
($39$) is much larger than for MWC~656 - like systems ($0.01$ -- it is a probability 
of having one MWC~656 - like system).
}
\label{Porb} 
\end{center}
\end{figure}

\begin{figure}
\begin{center}
\includegraphics[width=\columnwidth]{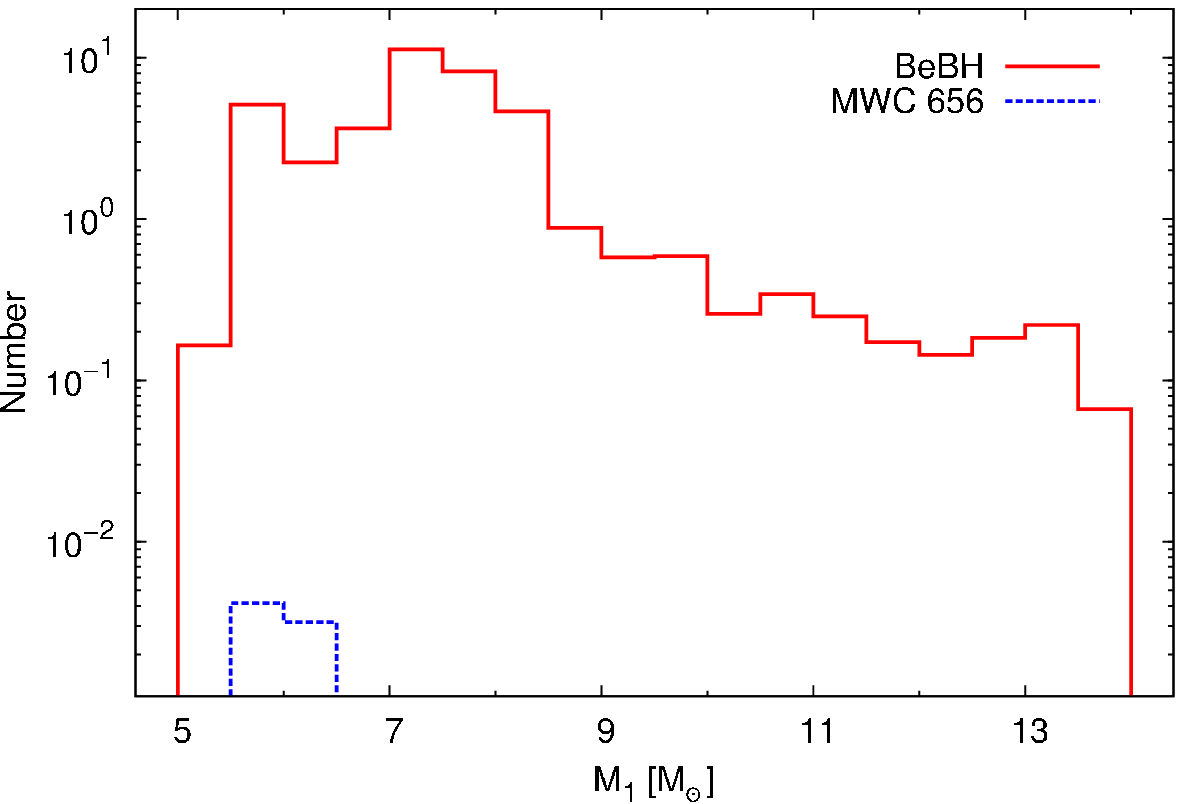}
\caption{Black hole mass distribution for overall group of B-BH
binaries (red solid line) and MWC~656 - like subpopulation (blue dashed line). 
Predicted current Galactic populations are shown.
}
\label{M1}
\end{center}
\end{figure}

\begin{figure}
\begin{center}
\includegraphics[width=\columnwidth]{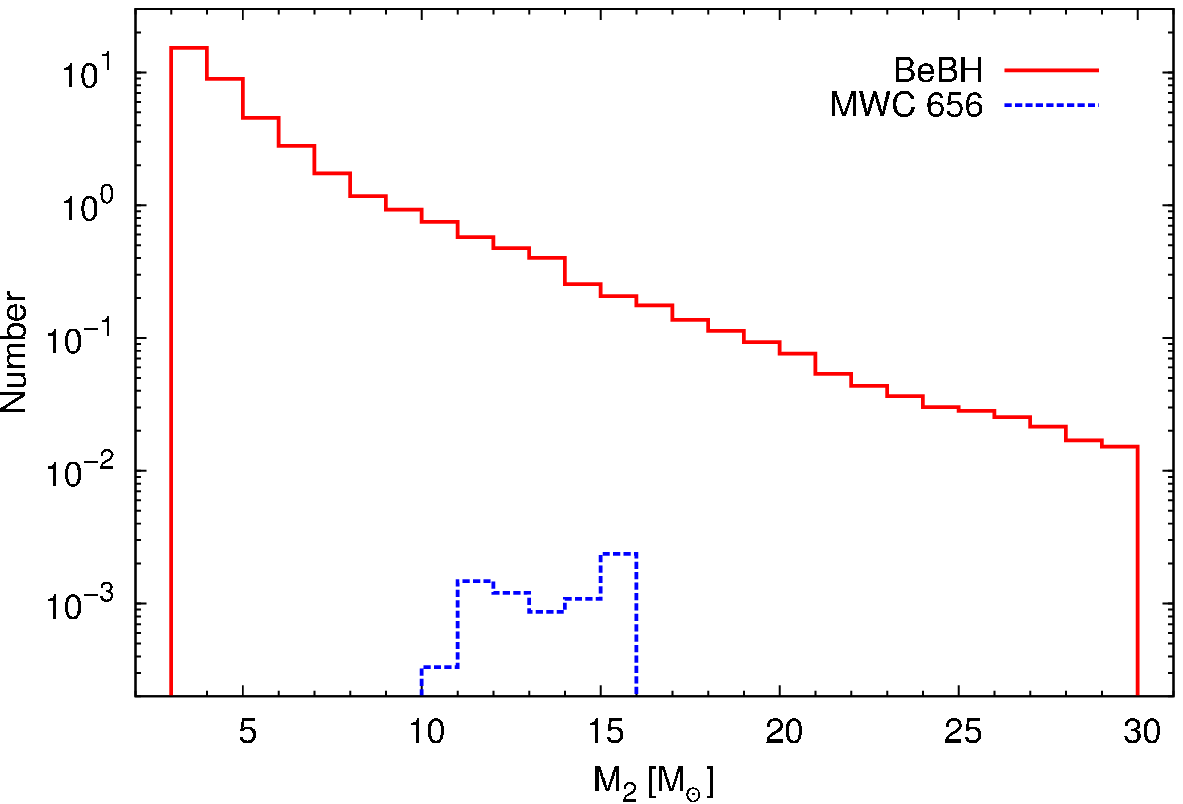}
\caption{B star mass distribution for overall group of B-BH binaries (red solid 
line) and MWC~656 - like subpopulation (blue dashed line). Predicted current Galactic 
populations are shown.
}
\label{M2}
\end{center}
\end{figure}

\begin{figure}
\begin{center}
\includegraphics[width=\columnwidth]{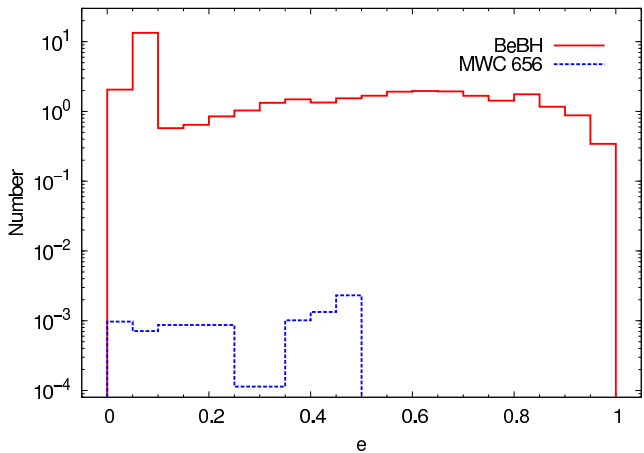}
\caption{Eccentricity distribution for overall group of B-BH binaries (red solid 
line) and MWC~656 - like subpopulation (blue dashed line). Predicted current Galactic 
populations are shown.
}
\label{e}
\end{center}
\end{figure}

\begin{figure}
\begin{center}
\includegraphics[width=\columnwidth]{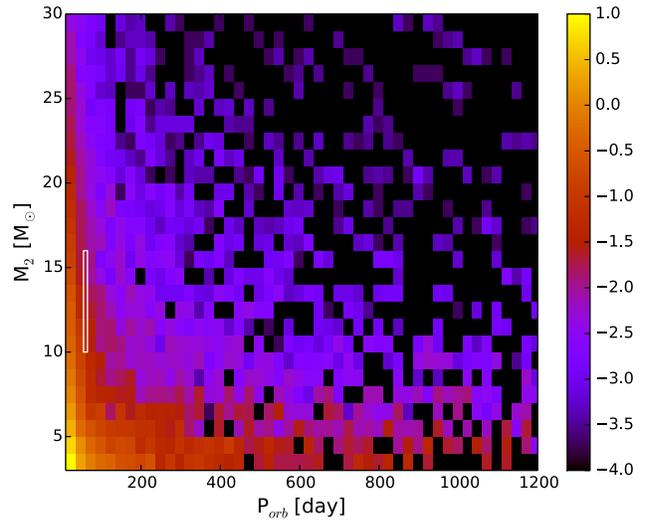}
\caption{
Two-dimensional distribution of orbital period ($x$ axis) and B star mass ($y$ axis)
for B-BH binaries. The numbers next to the color bar indicate the logarithm of the predicted 
current Galactic B-BH population.
The white rectangular box indicates the area of MWC~656-like systems defined in 
Eq.~\ref{MWC}.}
\label{P_M2}
\end{center}
\end{figure}

\begin{figure}
\begin{center}
\includegraphics[width=\columnwidth]{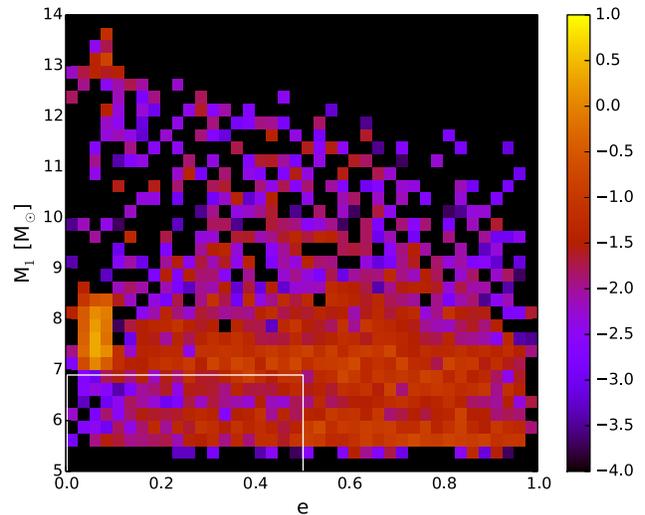}
\caption{
Two-dimensional distribution of eccentricity ($x$ axis) and BH mass ($y$ axis) for 
B-BH binaries. The numbers next to the color bar indicate the logarithm of the predicted 
current Galactic B-BH population.
The white rectangular box indicates the area of MWC~656-like systems defined 
in Eq.~\ref{MWC}.}
\label{e_M1}
\end{center}
\end{figure}

In Figures~\ref{Porb}, \ref{M1}, \ref{M2} and \ref{e} we show distributions of
orbital period, black hole mass, companion mass and orbital eccentricity for 
B-BH binaries (defined by Eq.~\ref{BeBH}). We also show a subpopulation of 
B-BH binaries that resemble the observed properties of MWC~656 (defined by 
Eq.~\ref{MWC}). 
In Figures~\ref{P_M2} and \ref{e_M1} we show the two-dimensional distributions 
of orbital period vs. B star mass and eccentricity vs. BH mass for B-BH binaries.
With a white rectangle we mark the region corresponding to systems defined as
MWC~656-like (Eq.~\ref{MWC}).

\begin{figure}
\begin{center}
\includegraphics[width=\columnwidth]{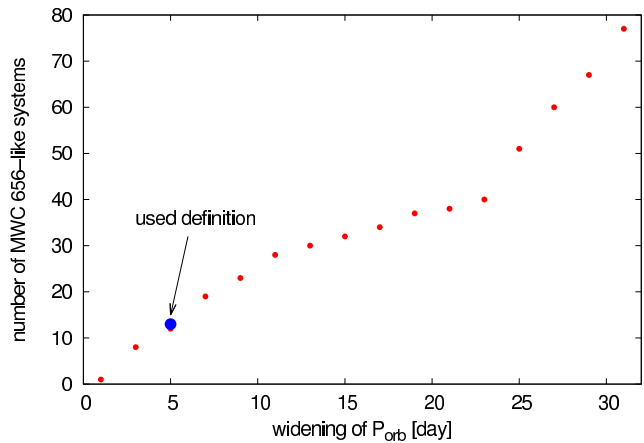}
\caption{
Sensitivity of formation efficiency of MWC~656-like systems (as defined in 
Eq.~\ref{MWC}). For our study we have adopted definition 
$55 < P_{\rm orb} < 65{\rm d}$ for MWC~656-like systems in respect to 
orbital period. For this definition we produce $13$ systems in entire 
$10$ Gyr of Galaxy evolution (see the blue dot). If we widen the period 
range by a given number of days (horizontal coordinate) on both {\em sides} of the 
measured orbital period of MWC~656 ($P_{\rm orb}=60.37$ d) then we obtain
specific number of MWC~656-like systems (vertical coordinate). }
\label{per_col}
\end{center}
\end{figure}

In Fig.~\ref{per_col} we present the sensitivity of total
number of MWC~656-like systems formed over entire $10$ Gyr in the Galaxy 
on the adopted orbital period range in the systems' definition Eq.~\ref{MWC}.

There is a two-stage evolution leading to the formation of B-BH binaries; CE 
followed by core-collapse/SN (forms BH). In the first stage the massive primary 
expands after MS and initiates RLOF. Due to the high mass ratio (typically 
$M_{\rm primary}/M_{\rm secondary} \gse 3$) mass transfer is unstable and 
leads to CE. At the onset of CE, our massive primary ($\sim 30$--$60\msun$) is
most likely ($\sim 90\%$) in the Core Helium burning (CHeB) phase and has a well 
developed convective envelope. Only in a relatively small number of cases 
($\sim 10\%$) the primary is on the Hertzsprung gap (HG) with either a radiative
or shallow convective envelope\footnotemark[2].
\footnotetext[2]{
Note that these numbers have changed significantly since \cite{Bel+Ziolk}, who 
reported only $\sim 4\%$ of progenitors of B-BH binaries to go through CE with
a donor on CHeB stage. This apparent discrepancy is a direct consequence of updates
of input physics made in $\tt StarTrack$ code. It mostly originates from
change of common envelope physics. In the updated code the i$\lambda$ parameter
that describes the binding energy of the donor star is calculated based on
stellar radius, metallicity and evolutionary stage. In the previous study
the authors used a constant value $\lambda=1$, which is now replaced with a
typical physical value of $\sim 0.1-0.2$ for BH progenitors on HG
(see Fig.~3 in \cite{Dominik2012}). It means, that now $\lambda$ is too
small to allow HG donor to survive CE, and these systems now are found to
typically merge during CE. The smaller lambda values preferentially select
donors with small envelope mass to survive CE. For massive BH progenitors
stars lose their H-rich envelopes during CHeB and that is what we note in
our new results.}
In the case of HG donor it is not at all clear
whether CE develops at all even for high mass ratios (unpublished MESA 
simulations). This $90$--$10\%$ division is an evolutionary selection effect. 
The secondary star is typically of much lower mass ($\sim 5$--$15\msun$; see 
Fig.~\ref{M2}) than primary. Therefore, it is easier for a binary to survive 
the CE phase for small CE envelope mass. HG stars have massive and tightly bound 
envelopes, while the envelope mass and binding energy decreases rapidly during
CHeB phase. Both as a result of expansion and for high metallicity also due to
intense wind mass loss. After CE initially wide systems evolve to much shorter
orbital periods ($\sim 10$--$100$ day; see Fig.~\ref{Porb}). Massive primaries
lose their entire H-rich envelope, and become Wolf-Rayet stars. We assume that
CE is very rapid and that the MS secondary does not have time to accrete. In 
our simulations we allow only compact objects (NS and BH) to accrete a small but
significant amount of mass during CE. Note that at the moment the CE phase is 
far from being understood \citep[see however][]{Ivanova2013} and our 
{\em assumptions} on accretion physics in CE are subject to verification.

In the second stage, the Wolf-Rayet star explodes in Type Ib/c SN and forms a BH. 
Alternatively, for very massive stars we assume direct BH formation without
accompanying SN. Our models for single stars allow for Wolf-Rayet star formation above 
$M_{\rm zams} \sim 20 \msun$ and for direct BH above $M_{\rm zams} \sim 40 \msun$. 
Roche lobe overflow episodes in binaries may significantly shift these boundaries 
up or down depending on a given binary configuration and evolutionary stage of 
its components. The associated with SN neutrino losses and potential mass 
loss and natal kick affect the orbit (altering eccentricity and the semi-major
axis). A BH binary with MS companion is formed. Such a binary may either fall 
right within our criteria for B-BH or MWC~656-like object directly, or it may 
evolve to satisfy these criteria at a later stage. The only process altering 
the binary orbit at this stage is wind mass loss from the MS star increasing 
the orbital separation (but this effect is very small for the MS stars in the 
mass range we consider). For close binaries tidal interactions (synchronizing 
MS stars and circularizing the orbit) also play a role. Since we have chosen 
our lower limit on the orbital period to be rather large (10 days), the systems 
that are subject to efficient tidal interactions are not typical progenitors 
of B-BH/MWC~656-like objects.

\subsection{The formation of MWC~656 - like system}
\label{past}

\begin{figure}
\begin{center}
\includegraphics[width=\columnwidth]{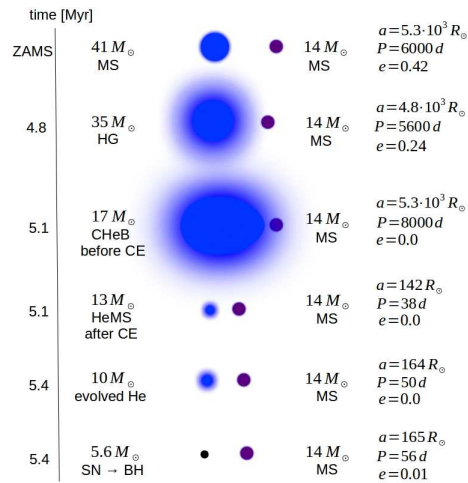}
\caption{
Typical evolution that may lead to the formation of MWC656 - like system. For the detailed  
description of the evolutionary history see Sec.~\ref{past}.
Note that the two most important evolutionary factors are 
common envelope phase that brings orbital period close to the current observed 
value and supernova Ib/c that forms rather light black hole similar to the one residing in MWC656.
\newline
The notation: 
ZAMS - zero age main sequence, MS - main sequence, HG - Hertzsprung gap, CHeB - core helium burning, 
HeMS - helium main sequence, BH - black hole, CE - common envelope, SN - super nova} 
\label{past_fig_2}
\end{center}
\end{figure}

Here, we present a typical evolutionary scenario that leads to the formation of 
a MWC~656 - like system (see Fig.~\ref{past_fig_2}).

We start the binary evolution with two components on the Zero Age Main Sequence 
(ZAMS) with $M_1 = 41 \msun$ (primary) and $M_2 = 14 \msun$ (secondary). The 
initial semi--major axis of the orbit is $a = 5.3 \cdot 10^3 \rsun$ and its 
eccentricity is $e = 0.42$ (the orbital period is $P_{\rm orb} = 6000$ day).
After $4.8$ Myr the primary with a mass $M_1 = 35 \msun$ finishes core 
hydrogen-burning and enters the HG. At this phase the primary significantly 
expands and tidal forces begin to circularize the orbit. Next, the primary 
enters CHeB expanding toward its Roche lobe. The orbit becomes fully 
circularized. This phase ends with the primary ($M_1 = 17 \msun$) overfilling 
its Roche lobe and initiating the common envelope phase. 
At $t=5.1$ Myr a close binary emerges out of the CE -- a $M_1 = 13 \msun$ 
helium core of the primary on a relatively close orbit (separation decreases 
from $a = 5300 \rsun$ to $a = 142 \rsun$ corresponding to decrease in orbital 
period from $P_{\rm orb} = 8000$ day to $P_{\rm orb} = 38$ day) around the 
mostly unaffected main sequence secondary $M_2 = 14 \msun$. 
After $\sim 0.35$ Myr the helium star primary finishes nuclear burning and its
mass decreases to $M_1 = 10 \msun$ due to strong Wolf-Rayet type winds. Just 
before the supernova (SN) explosion the semi--major axis of the orbit is 
$a = 167 \rsun$ and the orbital period $P_{\rm orb} = 52$ day (expansion due 
to the wind mass loss). The primary explodes in a type Ib/c SN (ejected mass 
$\sim 2.5\msun$) and forms a light black hole with mass $M_1 = 5.6 \msun$ (we 
have assumed $\sim 10\%$ mass loss in neutrino emission). We obtain a natal 
kick from an asymmetric mass ejection scenario and for this particular system the
magnitude of the kick is $130$ km\,s$^{-1}$ in such direction that it changes 
the separation to $a = 165 \rsun$ ($P_{\rm orb} = 56$ day) and eccentricity to 
$e = 0.01$. Depending on the orientation of the kick, the post-SN eccentricity
may vary in a broad range. For small post-SN eccentricities ($e<0.1$) systems 
have a likely chance to form BH--NS binary at the end of evolution, while more 
eccentric systems tend to finish evolution in CE mergers 
(see~Sec.~\ref{future}). 
Over the next $\sim 9$ Myr this relatively close binary remains almost 
unchanged with slight orbital expansion due to wind mass loss from the main 
sequence secondary ($a = 168\rsun$, $P_{\rm orb} = 58$ day). Throughout this 
phase the system is a wind-fed X-ray binary and it meets our criteria 
Eq.~\ref{MWC} and we tag it as the MWC~656-like system. 

Initially MWC~656 progenitors are two massive stars. If placed on a short orbit, 
such stars would begin interacting while on the MS or at the beginning of the HG 
and such a RLOF would most likely lead either to component merger \citep{Sana2012}
or would deplete the primary mass below the threshold of BH formation. For wide 
binaries, it takes a CE phase to decrease the orbital size to the currently 
observed period of MWC~656 (as described in our example). For very wide systems 
either CE is never encountered or if it is there is only very little mass in CHeB 
donor envelope and orbital decrease is not efficient enough to produce orbital 
period observed for MWC~656.

\subsection{Parameter study}
\label{sec_param}

Our results are subject to a number of evolutionary uncertainties. We have 
performed several additional calculations that probe the most obvious 
uncertainties in the formation of B-BH binaries. In models $V_1$ and $V_2$ we 
have altered the energy balance in CE evolution that is required in all channels 
leading to B-BH binary formation. In models $V_3$ and $V_4$ we have changed our 
treatment of BH formation (natal kicks and BH mass). These changes lead to 
different formation efficiency of B-BH binaries. In Table~\ref{number_binaries} 
we list total number of B-BH binaries formed over entire $10$ Gyr of Galactic 
evolution as well as their current predicted number in Galaxy for all models. 
The same numbers are listed for MWC~656-like systems. 
\begin{table*}
\begin{minipage}{110mm}
\caption{B star + black hole binary numbers in Milky Way$^a$}
\begin{tabular}{@{}cccrrccrrccl}
\hline
model &&& \multicolumn{2}{c}{MWC~656-like}  &&&  \multicolumn{2}{c}{B-BH} &&& comment     \\
\hline
$S_0$             &&&     13 & (0.007)    &&&  $ 8,700$ &  (39)  &&& standard evolution    \\
$V_1$             &&&      0 & (0)        &&&  $ 1,600$ &   (7)  &&& $\alpha_{\rm CE}=0.1$ \\
$V_2$             &&&     34 & (0.026)    &&&  $55,800$ & (131)  &&& $\alpha_{\rm CE}=5$   \\
$V_3$             &&&     71 & (0.063)    &&&  $12,800$ &  (63)  &&& no BH kicks           \\
$V_4$             &&&      8 & (0.004)    &&&  $13,300$ &  (50)  &&& delayed SN engine     \\
\hline
\end{tabular}
\label{number_binaries}
$^a_{\hspace{3mm}}$The total number of B-BH and MWC~656 - like systems formed in
simulations over $10$ Gyr of evolution of the Galaxy; in parenthesis
we list the current number of systems in the Galaxy.
\\
\end{minipage}
\end{table*}

The current number of B-BH systems is predicted at the level of $\sim 10$--$100$.
This number is rather sensitive to the adopted assumption on CE efficiency. 
For model in which we allow only $10\%$ ($\alpha_{CE}=0.1$; model $V_1$) 
of orbital energy to be used in CE ejection we predict only 7 B-BH systems 
to be currently present in our Galaxy. For much more efficient ejection, 
with $5$ times of orbital energy is used ($\alpha_{CE}=5.0$; model $V_2$) 
for CE ejection we find many more B-BH systems: $131$. We only allow 
orbital energy to increase in such arbitrary way as to mimic the possibility 
that bounding energy of CE is much lower than used in current predictions.
We use binding energy estimates from \cite{Xu2010} \citep{Xu2010b} that give 
binding energy of envelope for a given star radius, metallicity and star 
evolutionary stage. However, the envelope internal energy may lead to much 
easier ejection and it was estimated that it is potentially realistic to 
decrease binding energy by factor of $\sim 5$ \citep{Ivanova_2011}. That is 
what we have employed in model $V_2$. For our reference model we have used 
$100\%$ of orbital energy for CE ejection ($\alpha_{CE}=1.0$; model $S_0$). 

Typical B-BH formation starts with a rather massive star (BH progenitor) that 
forms massive envelope after main sequence. Ejection of massive CE ($10-30\msun$; 
at the end of Hertzsprung gap) by a typical B star ($\sim 5$--$15\msun$; see 
Fig.~\ref{M2}) is rather hard. In particular, for low ejection efficiency 
($\alpha_{CE}=0.1$) it leads to a decrease in number B-BH binaries. For model 
in which we allow for lowered binding energy (or high $\alpha_{CE}=5.0$) the 
B-BH number increases. Alternatively, binary channels are naturally selected 
in which the envelope of a massive primary is depleted by evolution (winds and 
core growth) and then late case C RLOF leads to CE development (with CHeB donor) 
and B-BH formation. 

The delayed SN model allows for the formation of low mass BHs (see \cite{Fryer2012}), 
as opposed to our standard model in which we do not allow formation of BHs in the mass 
gap: $3-5\msun$ \citep{Belczynski2012}. We note no significant change in number of 
B-BH binaries between these two models (standard vs. delayed SN). There is a small 
increase of B-BH binaries in the model with no BH kicks, since in this case some 
progenitors are not disrupted upon BH formation. Since the majority of BHs in B-BH 
binaries are predicted to have rather high mass (peak of mass distribution at $7-8 \msun$; 
see Fig.~\ref{M1}) then they receive small or no natal kicks in our standard model. 
Therefore, the change to zero BH kick has a small effect on the overall B-BH population. 
This is quite different for MWC~656-like systems, for which we note factor of $\sim 10$ 
increase of current Galactic number in model $V_3$. As low-mass BHs (as observed in 
MWC~656) receive non-zero kicks (binary disruptions) at the formation in our 
standard model, thus in model $V_3$ (no kicks) we note significant increase of
MWC~656-like systems in the overall population of B-BH binaries.

We note that MWC~656 eccentricity and peculiar space velocity are consistent 
with no or small natal kick. The high BH natal kicks (above $100$--$150$ km\,s$^{-1}$) 
may be excluded by the analysis presented in the Appendix.
In our particular example of the formation of MWC 656-like systems (presented 
in~Fig.\ref{past_fig_2}) the BH is formed with a moderate 3D natal kick $V_{kick}=130$ 
km\,s$^{-1}$ and mass loss of $\sim 3\msun$. This natal kick was oriented in such 
a way that eccentricity was essentially not affected ($e_{\rm postSN}=0.01$), but 
the systemic velocity was increased by $50$ km\,s$^{-1}$. This is close to the values
presented in the Appendix: $e<0.14$ and $V_{\rm spaceMWC656}<37.4$ km\,s$^{-1}$.
Within our population of MWC~656-like systems we find binaries that are fully 
consistent (in a 1-sigma range) with the Appendix estimates: {\em (i)} for natal kick 
of $23$ km\,s$^{-1}$ and mass loss of $\sim 3\msun$ we get $e=0.1$ and $V_{\rm spaceMWC656}=22$ 
km\,s$^{-1}$ and {\em (ii)} for natal kick of $80$ km\,s$^{-1}$ and mass loss of 
$\sim 3\msun$ we get $e=0.12$ and $V_{\rm spaceMWC656}=36$ km\,s$^{-1}$.  
The above examples were obtained with our standard model, which assumes
non-zero BH (but smaller than NS) kicks. Our model for non-zero kicks is
based on the asymmetric mass ejection from pre-supernova star. The amount of
mass loss (that sets the natal kick value) is based on the supernova models
presented in \cite{Fryer2012}. Apparently these models can explain both
the eccentricity and space velocity of MWC 656.
Obviously, our models with no BH natal kicks are also fully consistent with
the eccentricity and peculiar space velocity estimates. Therefore, we cannot
distinguish between the two models (asymmetric mass ejection versus no natal
BH kicks).

\subsection{On the origin of the Be phenomenon in Be-BH systems}

The emission lines of regular main sequence Be stars are generally considered 
to result from an outflowing disk of a star rotating at or at a substantial 
fraction of its Keplerian rotation rate or break-up velocity \citep{Rivinius2013}. 
This raises the question about the origin of the spin of the Be star. This may 
either be reflecting the birth spin of the star or it may be the consequence of 
interaction in the binary system.

The first case is plausible, since young massive stars are observed to rotate 
at a wide range of rotation rates \citep[e.g.][]{Abt2002, Ramirez2013, 
SimonDiaz2014}. It does however require the Be star to retain its spin through
all phases of binary interaction including the common envelope. This phase is 
poorly understood. Detailed stellar structure models show that rotating main 
sequence stars naturally tend to spin up their outer layers towards the 
Keplerian rotation rate. This does however require that the outer layers are 
well coupled with the contracting core. It also requires that angular momentum
loss in the form of stellar winds is not significant, which is the case for 
most B-type main sequence stars and at low metallicity for the later O type 
stars as well \citep[][]{Ekstrom2008b, Mink2013}. In the currently favored 
picture this can cause stars that are rotating sufficiently fast to reach 
break-up as they evolve and show the emission phenomenon until they leave the 
main sequence.

The second case requires a phase of mass transfer from the progenitor of the 
current black hole to the B star. \cite{Pols1991} showed that spin up by mass 
transfer is very efficient: accreting just a few percent of its mass from an 
accretion disk is sufficient to bring a star to break-up. Simulations by 
\cite{Pols1991} and \cite{Mink2013} show that, given the high binary frequency
among massive stars, this is expected to be a very important or possibly even 
dominant channel for the formation of rapidly rotating stars 
\citep[see also][]{ShaoLi2014}.

In our models we find that none of our MWC~656-like systems 
(containing both B and Be stars) formed through a 
formation channel that included a phase of stable Roche Lobe overflow. So our 
simulations give preference to the first case, where the spin of the B star is
the spin resulting from birth. However, it must be noted that in our model we 
rely on simplified prescriptions of the stellar structure and therefore an 
approximated treatment of the response of a star to Roche lobe overflow. One 
possibility is spin up during a (presumably short phase of) stable mass 
transfer preceding the CE. Another interesting alternative is stable mass 
transfer through atmospheric or wind Roche lobe overflow 
\citep[e.g.][]{Abate2013}.

\subsection{The future evolution of MWC~656-like system}
\label{future}

The fate of MWC~656-like system strongly depends on the secondary mass on 
ZAMS. We selected two broad evolutionary categories; binaries with relatively 
low-mass $10<M_{2_{ZAMS}}<13\msun$ and high-mass $13<M_{2_{ZAMS}}<16\msun$ 
secondaries. The summary of future evolution of MWC~656-like systems and the 
estimates of BH--NS formation chances are given in Table~\ref{evolution}.
\begin{table*}
\begin{minipage}{150mm}
\caption{Future evolution of MWC 656-like binaries$^a$}
\begin{tabular}{@{}crlcrrr}
\hline
 \multirow{2}{*}{Channel}          & \multirow{2}{*}{$f_{\rm form}$}       &  
\multirow{2}{*}{Evolutionary history$^b$}    & Mergers$^c$   &
\multicolumn{3}{c}{Fate$^d$ (BH--NS):} \\
        &          &                                       & CE/RLOF& Close    & Wide    & Disrupted \\
\hline
B-BH:1a & $15.4\%$ & CE1(4-1) SN1 MT2(14-2) MT2(14-9) ECSN2&  $0\%$   &  $0\%$   &$15.4\%$ & $0\%$    \\
B-BH:1b & $23.1\%$ & CE1(4-1) SN1 MT2(14-2) SN2            &  $0\%$   &  $0\%$   & $0.5\%$ & $22.6\%$ \\
        &          &                                       &        &          &         &          \\
B-BH:2a &  $7.7\%$ & CE1(4-1) SN1 CE2(14-4) MT2(14-7) SN2  &  $0\%$   &  $5.6\%$ & $0.7\%$ & $1.4\%$  \\
B-BH:2b & $53.8\%$ & CE1(4-1) SN1 CE2(14-2) MT2(14-7) SN2  & $38.4\%$ & $10.7\%$ & $1.3\%$ & $3.4\%$  \\
\hline
\end{tabular}
\label{evolution}
$^a_{\hspace{3mm}}$We list only formation channels of MWC 656-like systems which are defined by 
Eq.~\ref{MWC}.\\
$^b_{\hspace{3mm}}$Sequences of different evolutionary stages:
CE1 and CE2: common envelope with a primary and secondary as a donor, respectively; 
MT2: non--conservative mass transfer with a secondary as a donor; 
SN1 and SN2: type Ib/c supernova of the primary (black hole formation) and secondary
(neutron star formation), respectively; 
ECSN2: electron capture supernova of secondary (neutron star formation). 
\newline
Numbers in parenthesis denote evolutionary stage of primary--secondary: 
$1$ - main sequence, $2$ - Hertzsprung gap, $4$ - core helium burning,
$7$ - helium main sequence, $9$ - helium giant branch, $13$ - neutron star, $14$ - black hole.\\
$^c_{\hspace{3mm}}$This is probability that two binary components merge in RLOF or CE events
that are encountered between the two SNe events. \\ 
$^d_{\hspace{3mm}}$Outcome of future evolution of MWC 656-like systems; close (delay time
from ZAMS to BH and NS merger shorter than $10$ Gyr) or wide BH--NS systems or 
disrupted BH and NS objects may form. 
\end{minipage}
\end{table*}

\subsubsection{Progenitors of wide BH--NS binaries}

Here we describe, the future evolution of MWC~656-like systems with the 
initial (ZAMS) secondary mass smaller then $13\msun$. This group consists of 
$\sim 38\%$ of all MWC~656-like systems formed in Galactic disk, with $15.9\%$ 
forming wide BH--NS systems and $22.6\%$ are being disrupted in second SN. 

We start with the binary consisting of a BH with mass $M_1=5.7\msun$,  and a 
secondary star that has just entered HG with mass $M_2=10.2\msun$ (so right 
after MWC~656-like phase) and with separation $a=153\rsun$ 
($P_{\rm orb}=52.2$ day) and eccentricity $e=0.28$. 

The secondary quickly expands while crossing HG and it initiates stable Roche 
lobe overflow (RLOF) onto the BH. Rapid expansion does not allow for tidal 
circularization. We circularize the system instantaneously at the onset of RLOF, 
we take the periastron distance as the new separation ($a=110\rsun$, 
$P_{\rm orb}=34$ day) of circular orbit ($e=0$). The accretion onto BH is 
Eddington limited, and the rest of mass leaves the system with the BH specific
angular momentum. During RLOF the separation first decreases to $a=101\rsun$ 
($P_{\rm orb}=32$ day) and then increases to $a=364\rsun$ ($P_{\rm orb}=263$ 
day). The binary components go through a mass ratio reversal, with BH mass 
$M_1=7.1\msun$ and the secondary mass $M_2=2.3\msun$ at the end of RLOF. The 
secondary becomes a naked helium star with core Helium-burning that lasts 
about $4$ Myr. The low mass helium secondary begins to significantly expand after 
it becomes an evolved helium star (Helium shell burning) and it initiates another 
episode of RLOF. At this point, the secondary mass decreases to $M_2=2.1\msun$
and the binary separations increases to $a=423\rsun$ ($P_{\rm orb}=333$ day).
As a result of the RLOF the BH mass reaches $M_1=7.2 \msun$ and secondary mass
is depleted to $M_2=1.6\msun$, while the orbit expands to $a=623\rsun$ 
($P_{\rm orb} = 608$ day). The secondary ends its evolution as a neutron star 
(NS) of a mass $M_2 =1.26\msun$ created in electron capture SN. All binary 
systems survive the explosion as we assume no natal kick for electron capture 
SN and very little mass was lost in the process. This evolutionary channel is 
marked as ``B-BH:1a'' in Table~\ref{evolution}. We note the formation of wide 
BH--NS system with chirp mass of $M_c=2.4\msun$ and a very long merger time 
$t_{\rm merger}=3.9\times 10^8$ Gyr.

For slightly more massive secondaries ($M_{2_{ZAMS}}\approx12\msun$) than in 
the above case, the second RLOF episode is avoided and a NS with mass 
$M_2=1.1\msun$ is created in a type Ib/c SN explosion. The chance of survival 
of such a binary is only $\sim 2\%$ due to frequent natal kick disruptions. 
The systems that survive SN explosion form wide BH--NS binaries with merger 
times exceeding the Hubble time. This evolutionary channel is marked as 
``B-BH:1b'' in Table~\ref{evolution}.

\subsubsection{Progenitors of close BH--NS binaries}
\label{Sclose}

In this section we describe the fate of MWC~656-like system with a secondary 
mass on ZAMS larger than $13\msun$. Such systems make up $\sim 62\%$ of the 
population of MWC~656-like binaries formed in Galactic disk. Although a  
significant fraction of these systems merge in ensuing CE ($38.4\%$), and some
fraction gets either disrupted in the second supernova explosion ($4.8\%$) or form
wide BH--NS systems ($2.0\%$), a sizable fraction forms close BH--NS systems 
($16.3\%$). 

We start with the binary consisting of a BH with mass $M_1=5.7\msun$, and a 
secondary star that just has entered HG with mass $M_2=13.5\msun$ (right after
MWC~656-like phase) and with separation $a=168\rsun$ ($P_{\rm orb}=58$d) and 
eccentricity $e=0.01$. 

Typically, lower mass secondaries ($M_{2_{ZAMS}} \leq 14.5\msun$) evolve 
through HG and enter CHeB and then initiate CE (formation channel ``B-BH:2a'';
see Table~\ref{evolution}). The chances of survival of this CE are close to 
unity due to small CE envelope mass. Higher mass secondaries ($M_{2_{ZAMS}} 
\geq 14.5\msun$) initiate CE while still on HG (``B-BH:2b''). The chances of 
survival of this CE phase are only about one third due to large CE mass. It is
possible that instead of CE some of these systems evolve through fast (on 
thermal timescale) but stable RLOF. In such a case these channel will 
contribute very little (if any) to the formation of close BH--NS binaries. 

After CE, the secondary loses most of its mass and becomes a naked helium star 
with mass $M_2=3.3\msun$. The orbit is circularized and decreases in size to 
$a=1.5\rsun$ ($P_{\rm orb} = 1.7$ hour). A very compact binary is formed. The
core-Helium burning secondary expands and initiates a stable RLOF. After 
$\sim 2$ Myr the secondary becomes an evolved helium star. The RLOF continues. 
The BH mass increases to $M_1=6\msun$, while the secondary mass decreases to 
$M_2=1.6\msun$ and at the time of the SN explosion the orbit has expanded to 
$a=4\rsun$ ($P_{\rm orb} = 8.5$ hour). After $16.5$ Myr from the beginning of 
the evolution at ZAMS, the secondary explodes in SN type Ib/c, forming a NS 
of a mass $M_2=1.1\msun$. Chances for a binary survival are very high as these
systems are very compact. 

For one particular natal kick, the post-SN orbit becomes eccentric $e=0.46$ and 
the semi--major axis increases to $a=5.7\rsun$ ($P_{\rm orb}=14.4$ hour). The 
close BH--NS binary is formed with the chirp mass $M_{\rm c}=2.1\msun$ and 
merger time $t_{\rm merger}=1.5$ Gyr.

\subsection{Future evolution of MWC~656}
\label{future1}

\begin{figure}
\begin{center}
\includegraphics[width=\columnwidth]{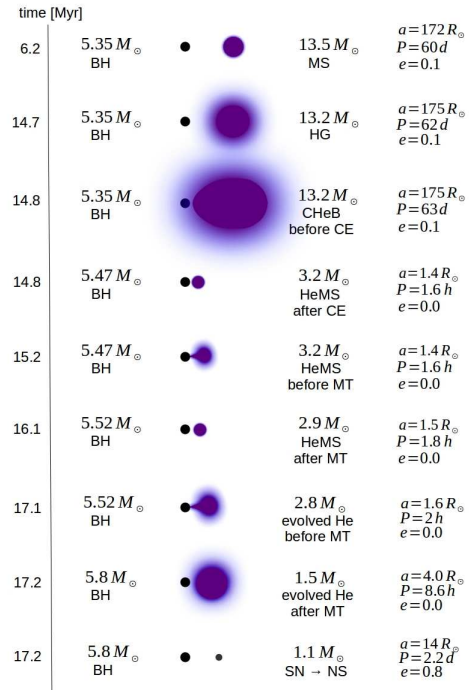}
\caption{ 
The future evolution of MWC 656 system (see Sec.~\ref{future1}). 
The binary evolves through CE and stable RLOF phase. If the binary survives CE phase the 
secondary star will form a light NS in supernova explosion ($1.1\msun$). 
Since CE significantly decreases the orbital separation, the binary is very likely to survive 
supernova mass loss and natal kick and to form a close BH--NS system (probability of $77\%$). 
\newline
We use the same notation as in Fig.\ref{past_fig_2}, with the addition of
MT - mass transfer, NS - neutron star.
} 
\label{future_fig_2}
\end{center}
\end{figure}

Based on the results presented in the above sections we can describe the future 
evolution of a binary that resembles MWC~656 (see Fig.~\ref{future_fig_2}). 
We start the evolution of the binary with a black hole mass $M_1=5.35\msun$, 
B star mass $M_2=13.5\msun$, semi--major axis $a=172\rsun$ and eccentricity 
$e=0.1$ and this results in an orbital period of $P_{\rm orb}=60.37$d \citep{Casares_2014}. 
The parameters are all within the observational errors of the original observational 
estimates and allow for the system to evolve along our evolutionary channel ``B-BH:2a''.
The massive primary ($M_{\rm zams}=30-35\msun$) took $\sim 6.2$ Myr to form a BH. 
A secondary MS lifetime is $\sim 14.8$ Myr. This gives us a typical lifetime of 
the B-BH binary phase of $t_{\rm MWC656}=8.6$ Myr. After finishing MS evolution, 
the secondary enters HG and starts to burn Helium in its core. During this stage 
a CE phase is initiated. The orbit significantly decreases after CE and the 
secondary becomes a low-mass naked helium star. The secondary expands again 
while core-helium burning and initiates stable Roche lobe overflow. A BH
increases its mass during CE (by $\sim 0.12 \msun$) and during stable RLOF (by
$\sim 0.3 \msun$) to $M_1=5.8\msun$. Finally, after $t_{\rm evol}=17.2$ Myr
from ZAMS the secondary explodes in type Ib/c supernova and forms a NS with 
mass $M_2=1.1\msun$. SN neutrino and mass loss ($\sim 0.32\msun$) and natal 
kick (drawn from Maxwellian with $\sigma=265$ km s$^{-1}$) may lead to a 
disruption of a binary (probability of $13.2\%$), formation of wide 
non-coalescing BH--NS system ($10.1\%$) and formation of close: coalescing 
within 10 Gyr BH--NS system ($76.7\%$). There is very high probability of close
BH--NS formation $f_{\rm close}=0.77$ as the circular pre-SN orbit is very 
compact and hard to disrupt ($a=4.0\rsun$; $P_{\rm orb}=8.6$h). We use a high 
number of {\em Monte Carlo} experiments to assess these probabilities and we 
show the resulting distribution of delay times in Fig.~\ref{tdelay}. The close
BH--NS systems that may form out of MWC~656 have average (and median) delay 
time of $1.9$ Gyr (and $0.9$ Gyr).

Note that initial stages of our prediction for future evolution of MWC~656
(Fig.\ref{future_fig_2}) resemble a recently discovered ULX source: P13 in NGC~7793 
\citep{Motch2014}. P13 is a binary with a $3-15 \msun$ BH and $\sim 18-23 \msun$ B9Ia 
companion. It was estimated that the system is experiencing the supercritical RLOF. 
The companion is proposed to be at the end of MS or at the beginning of HG. The orbital 
period is about 64 days. The only significant difference between P13 and MWC~656 is the 
mass of BH companion star. Since MWC~656 star has mass smaller than that in PG13, the RLOF 
will start later in its evolution. It is predicted that the onset of RLOF will begin when 
the star in MWC~656 will already finish crossing HG and will start Core He burning.
CE may be preceded by the short and stable high mass transfer rate phase (then
the MWC~656 would resemble P13). However, it seems that due to the existence of deep
convective envelope that forms during CHeB and quite high mass ratio (2.5) the development 
of CE is very likely.

Another object, SS433, was proposed to be a massive star engulfing a BH in its envelope 
\citep{Clark2007}. It may be the only known case of a massive binary undergoing CE. So far 
all the other binary mergers/CE events are restricted to low-mass stars (e.g., \cite{Kochanek2014}).

\subsection{Empirical LIGO/VIRGO detection rates}

\begin{figure}
\begin{center}
\includegraphics[width=\columnwidth]{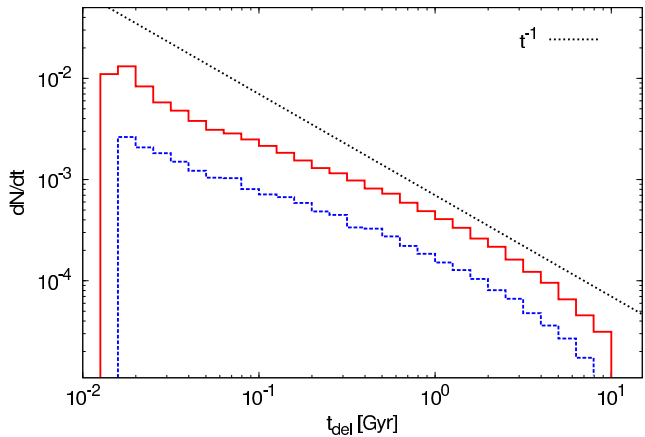}
\caption{
The delay time distribution for BH--NS systems formed out of  MWC~656-like population defined 
by Eq.~\ref{MWC} (red solid curve; average $1.8$ Gyr and median $0.8$ Gyr) and for BH--NS systems 
formed from the system with exact observed MWC~656 properties as defined in Sec.~\ref{future1} 
(blue dashed curve; average $2.0$ Gyr and median $1.1$ Gyr). 
}
\label{tdelay}
\end{center}
\end{figure}

The lifetime of MWC~656 may be estimated from our evolutionary calculations 
presented in Sec.~\ref{future1}; $t_{\rm MWC656}=8.6$ Myr. If we assume that 
only one such system is present currently in the Galaxy we obtain the Galactic
birth rate of ${\cal R}_{\rm birth} \approx 1/t_{\rm MWC656}$ under assumptions 
that star formation in Galaxy was constant and nothing special or extraordinary 
was required to form MWC~656. We have shown in Sec.~\ref{future1} that the 
probability of forming a close BH--NS binary out of MWC~656 is $f_{\rm close}=0.77$. 
Since the delay times are relatively short (median of the distribution is $0.9$ 
Gyr; see Fig.~\ref{tdelay}) as compared with the Galactic disk age ($10$ Gyr) we 
can estimate the Galactic merger rate as 
\begin{equation} \label{eq:rate}
{\cal R}_{\rm MW} = f_{\rm close} {\cal R}_{\rm birth} =  {f_{\rm close}
\over t_{\rm MWC656}}=0.089\ {\rm Myr}^{-1}
\end{equation}

We have converted the Galactic merger rates to the advanced LIGO/Virgo 
detection rates (${\cal R}_{\rm LIGO}$) assuming the constant density of Milky
Way-like galaxies at the level $\rho_{\rm gal} = 0.01$ Mpc$^{-3}$ in local 
Universe. We have adopted $d_0=450$ Mpc as the advanced LIGO/VIRGO horizon for
NS-NS binary (optimally oriented source with signal-to-noise ratio of 8) with 
chirp mass ${\cal M}_{\rm c,nsns} \equiv (M_1 M_2)^{3/5} (M_1+M_2)^{-1/5}=1.2 
\msun$ where individual NS masses are $M_1=M_2=1.4\msun$. The horizon for a 
double compact object with a given chirp mass ${\cal M}_{\rm c,dco}$ is 
calculated with $d=d_0({\cal M}_{\rm c,dco}/{\cal M}_{\rm c,nsns})^{5/6}$. 
Finally, the detection rate is obtained with:
\begin{equation}
{\cal R}_{\rm LIGO} = \rho_{\rm gal} { 4 \pi \over 3} \left({d_0 \over f_{\rm pos}}\right)^3 
\left( {\cal M}_{\rm c,dco} \over {\cal M}_{\rm c,nsns} \right)^{15/6} {\cal R}_{\rm MW}
\end{equation}
where factor $f_{\rm pos}=2.26$ takes into account the non-uniform pattern of 
detector sensitivity and random sky orientation of sources. For our case of BH
with $M_1=5.8\msun$ and NS with mass $M_2=1.1\msun$ we have 
${\cal M}_{\rm c,dco}=2.1\msun$ and the corresponding detection rate of 
${\cal R}_{\rm LIGO} = 0.115$ yr$^{-1}$. In other words, the existence of 
MWC~656 binary implies $1$ BH--NS advanced LIGO/Virgo detection in $9$ years. 

This estimate is subject to a number of uncertainties. For example, if we 
lower mass of the B star $\sim 10\msun$ then such a system will form only 
wide BH--NS binaries (see evolutionary channel ``B-BH:1a'' in 
Table~\ref{evolution}) and we get ${\cal R}_{\rm LIGO}=0$ yr$^{-1}$. If we 
lower B star mass to $\sim 11-12\msun$ such system will most likely get 
disrupted by core collapse SN (see evolutionary channel ``B-BH:1b'' in 
Table~\ref{evolution}) and we also obtain ${\cal R}_{\rm LIGO}=0$ yr$^{-1}$. 
These null rates are the direct result of our criterion on development of CE. 
For lower mass stars, instead of CE we encounter stable RLOF and therefore we 
form much wider systems that are either disrupted or do not merge within 
Hubble time. It is worth noting that the CE development and its inner 
mechanism is still far from being understood \citep{Ivanova2013}. On the other
hand we may increase the rate by increasing the B star mass to $\sim 15\msun$
(this shortens the B lifetime to $t_{\rm MWC656}=6.4$ Myr) and lowers the kick
velocities by factor of $2$ (this increases chance of SN survival to 
$f_{\rm close}=0.92$) to obtain ${\cal R}_{\rm LIGO}=0.187$ yr$^{-1}$ ($1$ 
detection in $5$ years). Lower NS kicks in interacting binary systems are 
supported for example by the observed ratio of single to binary millisecond 
pulsars \citep{Bel2010}.

\section{Discussion}
\label{discu}

We have studied the formation and future evolution of the first Be-BH binary 
system MWC~656. The study was carried out with population synthesis methods 
and it employed the standard model of single star and binary evolution.  

The formation of the system requires just two distinctive evolutionary steps. 
First, the massive primary (a BH progenitor) initiates a CE that is needed for
the formation of close binary with an orbital period similar to the one 
observed for MWC~656. Then, a Wolf-Rayet star (an exposed core of the primary)
explodes in type Ib/c SN and forms a $\sim 5\msun$ BH and makes the system 
eccentric, again as observed for MWC~656. At this point we note the formation 
of a massive binary consisting of a BH and a $10-16\msun$ B/Be star. 

Two competing scenarios are generally considered to explain the emission line 
phenomenon of Be stars. In both scenarios it is considered to be a direct 
result of rapid rotation and the presence of an outflowing disk. In the first 
scenario the B star is born as a substantially rapid rotator and retains this
spin throughout its MS evolution. In the case of our evolutionary calculations
it would also have to retain its spin throughout the CE phase initiated by its
companion. In the second scenario the B star was spin up through interaction 
with the progenitor of the black hole, most likely through accreting a small 
amount of mass. Just a few percent of mass gain would be sufficient.  

In our model we do not observe significant accretion and spin up of the secondary 
star during the CE phase, giving some support to the first scenario to explain the 
Be phenomenon in this system. However, our rapid binary code relies on simplified
prescriptions for stellar evolution and approximate treatment of the stars in 
response to mass transfer. It would be worth further investigating these 
formation channels in more detail with a full binary evolutionary code which 
properly solves the structure equations for both stars. Particularly interesting
promising possibilities for the formation of Be-BH systems include a  
phase of stable mass transfer (and thus spin-up by mass transfer) preceding the 
CE phases (delayed CE) or a phase of wind or atmospheric RLOF. 

At present the precise conditions for the Be phenomenon are not well understood.
With our models we cannot make definite conclusions about the fraction of B stars
that will show the phenomenon (and thus the fraction of B-BH binaries that appear
as Be-BH binaries).
In principle, explicit modeling of the stellar spins, the spin up process and the
moments of inertia will provide more information, for example as done by
\citet{Mink2013} and \citet{ShaoLi2014} under the assumption of rigid body rotation.
However, such simulations will not constrain the many remaining uncertainties.
Among the most important unknowns is the mass transfer and the resulting spin up
during accretion. It is still not well understood how the accretion stream interacts
with the star or the accretion disk and how the accreting star responds when its outer 
layers are spun up. Another important ingredient is the transfer of angular momentum 
throughout the star. Detailed stellar evolutionary models with and without
interior angular momentum transport by magnetic fields give different results
\citep[e.g.][]{Brott+2011, Ekstrom+2012}.
The situation is even less clear in the case of CE evolution. Particularly, it is unclear
whether a star can accrete mass and spin up before or during the CE phase.
\citep{MacLeod2015} find that the inspiraling object may increase its mass by a few percent.
This may be sufficient to spin up the accreting star \citep{Packet1981}.
As a general caveat the CE hydrodynamical simulations are still at the early stage
and do not reproduce the observations \citep{Passy2012}.
Finally, the initial distribution of stellar rotation rates is not well constrained.
The observations for young massive stars show a bimodal distribution with most stars
rotating with slow or moderate rates with a tail of rapid rotators reaching breakup
velocity \citep[e.g.][and references therein]{Dufton+2011, Dufton+2013, Ramirez2013, 
SimonDiaz2014}.
Considering all these uncertainties we opted for not following the spin evolution
explicitly in this study. We investigated the formation channels of 
B-BH systems. A certain fraction of these will (intermittently) show the Be phenomenon.
This fraction is uncertain. It may be as low as 0.3 if the ratio of Be stars to B
stars is comparable to that observed in clusters. It may be higher if the interaction
with the progenitor of the BH caused the Be phenomenon. A detailed 
investigation of the precise conditions for the occurrence of the Be phenomenon
could prove to be very valuable.

Although it was found that several tens of B-BH binaries are currently 
residing in the Galactic disk, we find that there is only small probability 
($\sim 1/100$) of having system resembling MWC~656 in terms of component 
masses and orbital period. In particular, in our simulations we find more 
systems with shorter orbital periods ($\sim 5-10$d), less massive B stars 
($\sim 3-5\msun$) and slightly more massive BHs ($\sim 7-8\msun$). For 
comparison, MWC~656 has orbital period of $\sim 60$d, Be star mass of 
$10-16\msun$ and BH mass of $\sim 5\msun$. If MWC~656 is representative of 
the intrinsic Galactic population of Be-BH binaries, it means that either our 
employed evolutionary model of massive stars or our adopted approach to CE 
and/or BH formation needs to be revised. We have employed non-rotating stellar
models from \cite{Hurley2000} with stellar winds corrected for clumping from 
\cite{Vink2001}. If a massive primary happens to be a fast rotator then our 
non-rotating models may not be a good choice, as one would expect smaller 
radii and larger cores for rapidly spinning stars. This would affect the 
development and the outcome of the CE phase and would also most likely lead to
the formation of more massive BH (\cite{Leitherer2014}, \cite{Mink2013}). The 
decreased stellar wind mass loss rates that we use are typically adopted in 
most recent evolutionary studies of massive stars. However, it appears that 
there may be some observational evidence for higher wind mass loss rates from 
massive stars \citep{Eldridge2008}. Had we adopted stronger winds, the primary 
core mass would decrease and to some extent it would counter-balance the 
evolutionary effects of fast rotation with the formation of a more massive BH.
For CE evolution we have adopted the energy balance model of 
\cite{Webbink1984} updated with the physical estimates of primary binding 
energy \citep{Xu2010, Xu2010b}. It appears that earlier claims that the core 
definition may change the post-CE binary separation by almost $\sim 2$ orders 
of magnitude \citep{DewiTauris2001} for stars with $M<20\msun$ (NS 
progenitors) does not apply to more massive stars (i.e., BH progenitors; 
\cite{Wong2013}). However, it is not at all clear that the energy balance 
model is good approximation for CE evolution, but no better model exists at 
the moment \citep{Ivanova2013}. For the BH formation we employ the rapid 
supernova model from \cite{Fryer2012}. This model explains observed mass gap 
between NSs and BHs \citep{BelczynskiWiktorowicz2012}, the lack of compact 
objects in $2$--$5\msun$ mass range. Our model also allows us to reproduce the 
Galactic and extragalactic BH mass spectrum \citep{BelczynskiBulik2012}. It is
possible that the mass gap is caused by some observational bias involved in BH
mass measurements as proposed by \cite{Kreidberg2012}. However, we have shown 
that our results do not depend sensitively on this aspect of evolution (i.e., 
BH formation mass; see model $V_4$). We have adopted an asymmetric mass ejection 
mechanism for BH natal kicks and this results in small BH kicks (decreased 
with the amount of fall back). The empirically derived kick velocities for 
$14$ Galactic BH binaries are inconclusive and the observational data allows 
for both small and high BH natal kicks \citep{Belczynski2012}. Had we adopted 
high BH natal kicks (similar to those measured for single pulsars; e.g., 
\cite{Hobbs2005}) the formation rates of B-BH binaries would decrease by 
large factor ($\sim 1$--$2$ orders of magnitude). In such the case we would 
expect very few Be-BH binaries to reside currently in the Galaxy. We have also
tested the alternative model with BH natal kicks all set to zero. For this 
model we note a significant (a factor of $\sim 10$; model $V_3$) increase of 
MWC~656-like systems currently predicted to reside in Galaxy. 

With the recent discovery of a Oe star with a compact companion \citep{Clark_2015}
we can await more Be-BH binaries to be found.
When more Be-BH binaries are identified it will be essential to measure the 
intrinsic distribution of their orbital periods. Matching the intrinsic 
distribution with evolutionary models may allow us to improve our understanding 
of BH natal kicks and may offer some insights into the inner workings of CE, as
these processes seem to be the most important factors affecting the formation 
of MWC~656-like systems and B-BH binaries in general. 

The future evolution of MWC~656 offers a very interesting potential of 
forming a BH--NS system. In particular, close BH--NS systems are a class of 
gravitational-radiation sources for Advanced LIGO \citep{AdvLIGO} and Virgo 
\citep{AdvVirgo} detectors, that are expected to start operation in a few 
years. We have estimated the formation rate of close (delay time from 
formation on ZAMS to BH--NS coalescence shorter than $10$ Gyr) BH--NS systems 
from binaries similar to that of MWC~656. We have translated the formation 
rate to detection rate of BH--NS mergers by advanced LIGO/Virgo. The advanced 
LIGO/Virgo detection rate is found to be up to $1$ detection in $5$ years.

This empirically inferred detection rate is comparable to the rate obtained 
from analysis of Cyg X-3 binary ($1$ detection in $10$ years; \cite{CygX3}) 
and is much higher than obtained for Cyg X-1 ($1$ detection in $100$ years; 
\cite{CygX1}). These empirical estimates are based on existence of some 
particular binary stars and therefore they are only lower limits (i.e., 
formation of BH--NS only along one very specific formation channel) to the 
detection rate. However, it is encouraging that the observational evidence, 
although indirect, increases to support the existence of close BH--NS systems. 
It is now the total of three systems: Cyg X-1, Cyg X-3 and MWC~656 that were 
shown to be potential BH--NS progenitors. Other known high-mass X-ray binaries 
were analyzed and excluded as potential BH--NS progenitors \citep{BelBul2012}.
Recent population synthesis analysis of overall formation channels of double 
compact objects further supports the empirical evidence with estimates of 
BH--NS merger advanced LIGO/Virgo detection rates at the level $0.03-5.7$ 
yr$^{-1}$ (from $1$ in $30$ years to $6$ per year; \cite{Dominik3}). Broader 
detection range is reported by \cite{Mennekens2014}: $0.04-484$ yr$^{-1}$ 
(from $1$ in $25$ years to $484$ per year). So far there are no known 
stellar-origin BH--NS binaries. At the moment the only known potential BH--NS 
binary consists of a central Galactic supermassive BH ({\em Sagittarius 
A$^\star$}) and its nearby magnetar PSR J1745-−2900 \citep{Eatough2013}.

\section{Conclusion}
At present we know around 180 X--ray binaries \citep{Ziolkowski2014}. For $\sim 120$
of them the nature of compact object was confirmed to be a neutron star. Just in
2014 the first Be X--ray binary with a black hole was discovered (MWC~656; \citep{Casares_2014}).

In this study we have investigated the possible evolutionary scenarios leading to the
formation of B-BH systems, part of which are Be-BH binaries. 
It was found that the B-BH progenitors experience common
envelope phase. For the majority ($90 \%$) of cases the CE donor is already an evolved
star (CHeB), which increases chances of envelope ejection and a survival of the binary.
In our simulations we used a code based on a non-rotating stellar models.
Therefore, we are not able to resolve the issue of the origin of rapid rotation of Be
stars. In our models it may be either connected to high initial star rotation or to
spin up in CE (or pre-CE mass transfer) phase.

We checked the sensitivity of our results on the various evolutionary parameters assumed
in our simulations. For all presented models we expect up to few tens of B-BH systems to reside in
the Galaxy at any given time, among which around 1/3 contain a Be star,
but the chance to have a system very similar to the MWC~656
is less than a few percent. The future discoveries of Be-BH systems may allow us to
determine the intrinsic distribution of orbital parameters and, by matching them with
models, to improve our understanding of phases like CE and SN, which play an important
role in the evolution of not only Be-BH binaries, but binaries in general.

We investigated the fate of systems similar to MWC~656. The future evolution of such
systems may lead to the formation of BH--NS binaries. We find that $18 \%$ ($7\%$)
of such population will form close (wide) BH-NS systems, while the rest merges in the
second CE phase or is disrupted in the second SN. We have repeated the prediction for
a system exactly like MWC656. Due to its favorable configuration, MWC656 is very likely
($\sim 77\%$) to form a close BH-NS system that will merger within 10 Gyr.

These results make MWC~656 along with Cyg~X-1 and Cyg~X-3 the only reported
potential progenitors of BH--NS binaries. The existence of MWC656 alone implies that
the detection of BH-NS mergers by advanced LIGO/VIRGO can be as high as 1 in
every 5 years.

\section*{Acknowledgments}
The authors acknowledge the Texas Advanced Computing Center (TACC) at The 
University of Texas at Austin for providing computational resources. 
MG and KB acknowledge support from Polish Science Foundation "Master2013" Subsidy.
MG acknowledge support by Polish NCN grant Preludium (2012/07/N/ST9/04184).
The work of KB was supported by the NCN grant Sonata Bis 2 (DEC-2012/07/E/ST9/01360),
the National Science Foundation under grant No. PHYS-1066293 and the hospitality of 
the Aspen Center for Physics. 
JC acknowledges support by DGI of the Spanish Ministerio de Educaci\'on, Cultura y 
Deporte under grants AYA2010-18080, AYA2013-42627 and SEV-2011-0187."
SdM acknowledges support for early stages of this study by NASA through an
Einstein Fellowship grant, PF3-140105 and a Marie Sklodowska-Curie
Reintegration Fellowship (H2020-MSCA-IF-2014, project id 661502).
The work of IN and AH is partially supported by the Spanish Ministerio de 
Econom\'{\i}a y Competitividad (MINECO) under grant AYA2012-39364-C02-01/02, 
and the European Union.
MR acknowledges support by the Spanish MINECO under grant FPA2013-48381-C6-6-P.
IR acknowledges support from the Spanish Ministry of Economy and Competitiveness 
(MINECO) and the Fondo Europeo de Desarrollo Regional (FEDER) through grant 
ESP2013-48391-C4-1-R.
JMP acknowledges support by the Spanish MINECO under grant AYA2013-47447-C3-1-P 
and financial support from ICREA Academia.
MB acknowledges support from the National Science Foundation under 
award HRD-1242090 and the hospitality of the Aspen Center for Physics.

\bibliographystyle{mn2e}
\bibliography{bibliografia}

\appendix
\section{Constraints on the SN explosion in MWC~656}
In this Appendix we report on the space velocity of MWC 656 and on constraints 
on the SN explosion that originated the BH in this system.

The position of MWC~656 in the ICRS (epoque J2000) according to \cite{Leeuwen2007} is:
\begin{equation}
\begin{gathered}
\alpha =  (22\, 42\, 57.30295 \pm 0.67) \textrm{\, mas} \\
\delta =  (+44\, 43\, 18.2525 \pm 0.72) \textrm{\, mas}
\end{gathered}
\end{equation}
where $\alpha$ and $\delta$ means right ascension and declination, respectively.
Its proper motion from the same reference is:
\begin{equation}
\begin{gathered}
\mu_{\alpha} \cdot \cos(\delta) = (-3.56 \pm 0.72) \textrm{\, mas$\, $yr$^{-1}$} \\
\mu_{\delta}                     = (-4.05 \pm 0.76) \textrm{\, mas$\, $yr$^{-1}$}
\end{gathered}
\end{equation}
From \cite{Casares_2014} the radial velocity of MWC~656 is ($-14.1 \pm 2.1$) km$\, $s$^{-1}$
and its distance is ($2.6 \pm 0.6$) kpc.

The Galactic rotation curve of Model A5 in Table 4 of \cite{Reid2014} provides:
\begin{equation}
\begin{gathered}
\rsun          = (8.34 \pm 0.16)  \textrm{\, kpc}  \\
\Theta_{\odot} = (240 \pm 8)      \textrm{\, km$\, $s$^{-1}$} \\
d_{\Theta_{\odot}}/d_{R_{\odot}} = (-0.2 \pm 0.4)   \textrm{\, km$\, $s$^{-1}$$\, $kpc$^{-1}$}
\end{gathered}
\end{equation}
The same model provides a peculiar space velocity of the Sun relative to
the Local Standard of Rest of:
\begin{equation}
\begin{gathered}
U_{\odot} = (10.7 \pm 1.8) \textrm{\, km$\, $s$^{-1}$}  \\
V_{\odot} = (15.6 \pm 6.8) \textrm{\, km$\, $s$^{-1}$}  \\
W_{\odot} =  (8.9 \pm 0.9) \textrm{\, km$\, $s$^{-1}$}
\end{gathered}
\end{equation}
The coordinates adopted for the North Galactic Pole and Zero longitude are
those listed in the Appendix of \cite{ReidBr2004}:
\begin{equation}
\begin{gathered}
\alpha_{\rm pole}  =  12^h\, 51^m\, 26.282^s = 192.8595083 \textrm{\, deg} \\
\delta_{\rm pole}   = +27^{o}\, 07^{'}\, 42.01^{''}  = 27.12833611 \textrm{\, deg} \\
\theta = 122.932 \textrm{\, deg}
\end{gathered}
\end{equation}

Using all these values we obtain a peculiar space velocity of MWC~656:
\begin{equation}
\begin{gathered}
v_{\rm space_{MWC656}} = (22.5 \pm 14.9) \textrm{\, km$\, $s$^{-1}$}
\end{gathered}
\label{vsp}
\end{equation}
The uncertainty was calculated using values listed in Tab.~\ref{uncert}, 
which added in quadrature give 14.9 km$\, $s$^{-1}$. Therefore, the measurement 
of space velocity is only significant at the 1.5 sigma level. Even for a fixed 
distance with no uncertainty the total uncertainty is 12.7 km$\, $s$^{-1}$, 
providing not even a 2-sigma detection. To obtain a 3-sigma detection it is 
required to provide null uncertainties for both the distance and the galactic 
rotation curve.

\begin{table}
\caption{The contribution of individual uncertainty to the space velocity uncertainty
of MWC~656.}
\begin{tabular}{@{}lc}
\hline
Considered uncertainty & Measured value (km$\, $s$^{-1}$) \\
\hline
Proper motion                   & 5.2 \\
Radial velocity                 & 1.7 \\ 
Distance                        & 7.8 \\
Galactocentric distance         & 0.5 \\
Galactic rotation of the Sun    & 6.8 \\ 
Galactic rotation around MWC656 & 7.0 \\
Peculiar space velocity of Sun  & 5.8 \\ 
\hline
\end{tabular}
\label{uncert}
\end{table}

The masses and orbital parameters of MWC~656 from \cite{Casares_2014} are:
\begin{equation}
\begin{gathered}
M_1 = (13 \pm 3) \msun \\
M_2 = (5.3 \pm 1.5) \msun \\
P = (60.37 \pm 0.04) \textrm{d} \\
e = 0.10 \pm 0.04
\end{gathered}
\end{equation}

Considering the large orbital period and the small eccentricity it is
reasonable to assume that the current value of eccentricity is similar to the one
just after the SN explosion, $e_{\rm postSN}$. In such a case, the mass lost
during the SN explosion is similar or slightly above to:
\begin{equation}
\begin{gathered}
\Delta_M = e_{\rm postSN} (M_1 \cdot M_2) = (1.8 \pm 0.8) \msun 
\end{gathered}
\end{equation}
Using the formalism described in \cite{Nelemans1999} (although the
orbital period here is much larger than 7 days) we find reduced mass, 
re-circularized period and initial orbital period:
\begin{equation}
\begin{gathered}
\mu = 0.91 \pm 0.03                 \\
P_{\rm recirc} = (59.5 \pm 0.7) \textrm{d}      \\
P_{\rm init} = (49 \pm 4) \textrm{d}
\end{gathered}
\end{equation}
With all these parameters we obtain an expected MWC~656 space velocity of:
\begin{equation}
\begin{gathered}
v_{\rm exp_{MWC656}} = (10.2 \pm 4.7) \textrm{km$\, $s$^{-1}$}
\end{gathered}
\end{equation}
to be compared to the measured space velocity of ($22.5 \pm 14.9$) km$\, $s$^{-1}$
from Eq.~\ref{vsp}.

Both values are compatible at the 1 sigma level, indicating that there is
no need of any additional kick to produce the observed space velocity with
a symmetric SN explosion with a mass loss of ($1.8 \pm 0.8$) $\msun$.

If we use the Eq.7 from \cite{Nelemans1999} to compute the
mass lost in the SN explosion considering the measured space velocity and
all other values with their corresponding uncertainties, we obtain:
\begin{equation}
\begin{gathered}
\Delta_M = (4.0 \pm 3.4) \msun 
\end{gathered}
\end{equation}
where the uncertainty was calculated using values from Table~\ref{uncert2}.
As expected, this is compatible with the value derived from the current
masses and the eccentricity of the orbit.

\begin{table}
\caption{The contribution of individual uncertainty to the uncertainty of the mass lost in the SN explosion of MWC~656.}
\begin{tabular}{@{}lc}
\hline
Considered uncertainty & Measured value ($\msun$) \\
\hline
Space velocity            &      2.6 \\ 
Mass of Be star           &      2.0 \\
Mass of BH                &      0.6 \\
Re-circularized period    &      0.02 \\
\hline
\end{tabular}
\label{uncert2}
\end{table}

In conclusion, everything seems compatible with no kick and a moderate
mass loss of a few solar masses to produce both the observed eccentricity
and the space velocity of MWC 656.

\end{document}